\documentclass{article}

\usepackage{arxiv}

\usepackage[utf8]{inputenc} % allow utf-8 input
\usepackage[T1]{fontenc}    % use 8-bit T1 fonts
\usepackage{hyperref}       % hyperlinks
\usepackage{url}            % simple URL typesetting
\usepackage{booktabs}       % professional-quality tables
\usepackage{amsfonts}       % blackboard math symbols
\usepackage{nicefrac}       % compact symbols for 1/2, etc.
\usepackage{microtype}      % microtypography
\usepackage{lipsum}		% Can be removed after putting your text content
\usepackage{graphicx}
\usepackage[numbers]{natbib}
\usepackage{doi}
\usepackage{amsmath}

\title{Multispectral Imaging with Fresnel Lens}

\author{%
    \textbf{Khen Cohen}\thanks{Corresponding author} \\
    Department of Electrical Engineering\\
    Tel Aviv University\\
    Ramat Aviv, Tel Aviv 69978, Israel \\
    \texttt{khencohen@mail.tau.ac.il} \\
    \And
    \textbf{Tuval Kay} \\
    Department of Digital Science\\
    Tel Aviv University\\
    Ramat Aviv, Tel Aviv 69978, Israel \\
    \texttt{tuvalkay@mail.tau.ac.il} \\
}

\date{}

\renewcommand{\headeright}{}
\renewcommand{\undertitle}{}
\renewcommand{\shorttitle}{\textit{arXiv} Template}

\begin{document}
\maketitle

\begin{abstract}
This paper presents a Multispectral imaging (MSI) approach that combines the use of a diffractive optical element, and a deep learning algorithm for spectral reconstruction. Traditional MSI techniques often face challenges such as high costs, compromised spatial or spectral resolution, or prolonged acquisition times. In contrast, our methodology uses a single diffractive lens, a grayscale sensor, and an optical motor to capture the Multispectral image without sacrificing spatial resolution, however with some temporal domain redundancy. Through an experimental demonstration, we show how we can reconstruct up to 50 spectral channel images using diffraction physical theory and a UNet-based deep learning algorithm. This approach holds promise for a cost-effective, compact MSI camera that could be feasibly integrated into mobile devices.
\end{abstract}

\section{Introduction}
Multispectral imaging is a non-invasive and innovative technology that captures image data within specific wavelength ranges across the electromagnetic spectrum, allowing the extraction of hidden or otherwise unseen information, finding valuable applications in numerous fields. This includes, but is not limited to, biomedical diagnosis \cite{liu2016application}, dermatological assessment \cite{ilisanu2023Multispectral}, agriculture inspection \cite{hassan2019rapid}, and food quality and safety control\cite{article,health2}.

Multispectral imaging (MSI) techniques can be segmented into several categories, including: scanning-based imaging, snapshot imaging, compressive sensing, and diffractive optics-based imaging. Scanning-based imaging involves scanning the target scene with a narrow-band filter, which is accurate but time-consuming \cite{Li2020Sampling}. Snapshot imaging captures multiple spectral bands simultaneously but may sacrifice spatial resolution and signal-to-noise ratio \cite{snapshot}. Compressive sensing reconstructs the full datacube from a small number of measurements, but it requires increased computational complexity and does not always detect high frequencies in the signal \cite{arce1999compressive}. Diffractive optics-based imaging splits light into its constituent wavelengths and captures them with a sensor array, providing high spatial resolution and spectral fidelity at a higher cost \cite{mengu2023diffractive}. The choice of MSI method depends on the specific application and system prerequisites.

A Fresnel lens is a uniquely constructed lens, light in weight and thin in form. It features an array of concentric rings carved into its surface, enabling the lens to concentrate light similarly to a traditional lens. However, it's critical to note that the focusing of light here occurs predominantly through diffraction rather than refraction, leading to a pronounced wavelength-dependence for the lens's focusing distance.

\begin{figure}[htbp]
\centering\includegraphics[width=1\textwidth]{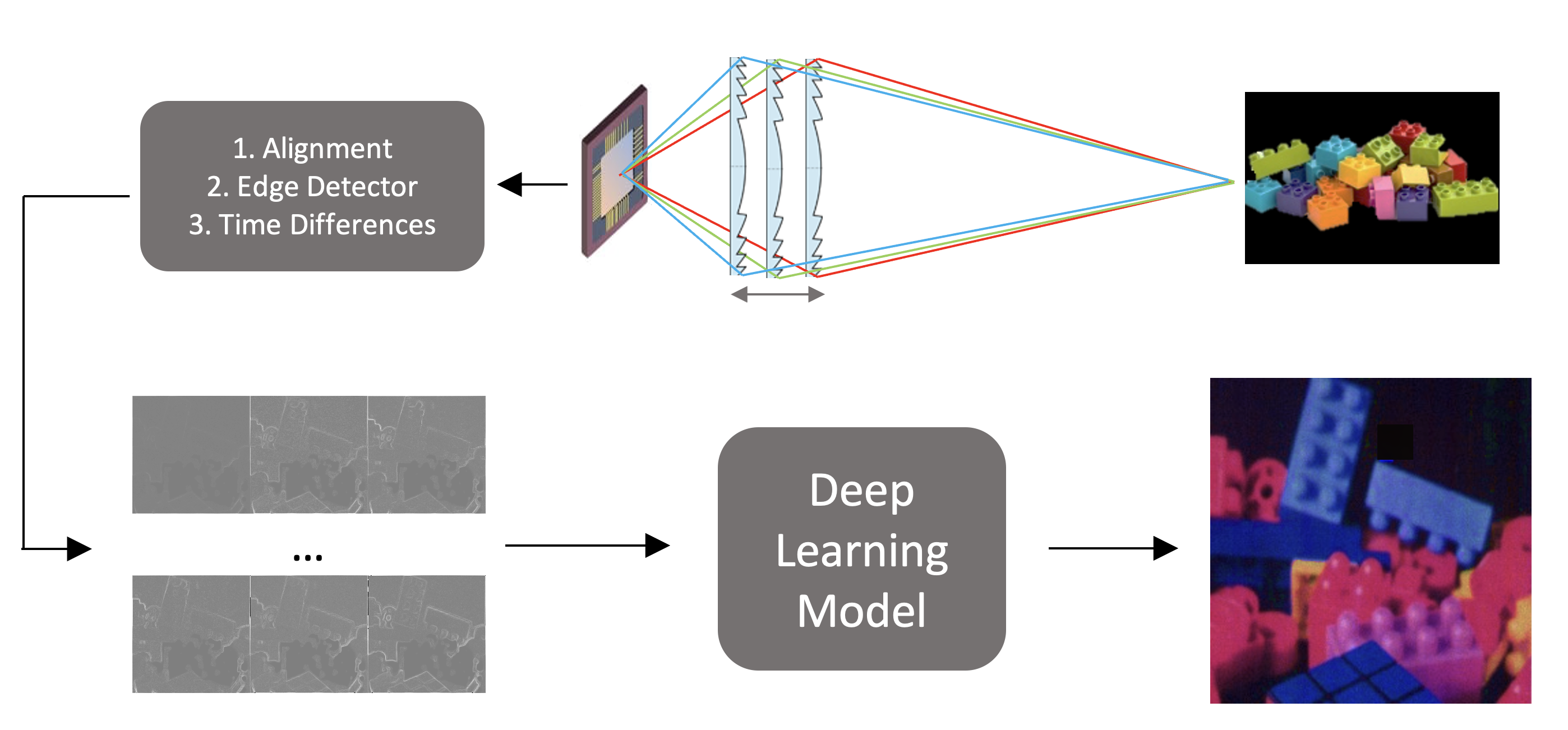}
\caption{A diagram of our method. We first capture a series of grayscale images at various lens positions. Subsequently, we process this data and employ a Deep Learning model to reconstruct the Multispectral image.}
\label{setup_diagram}
\end{figure}

While there's an unavoidable trade-off between system complexity (and cost) and image resolution and accuracy, our approach presents a method to achieve high resolution without compromising the system's complexity. Our proposed system does exhibit a limitation when it comes to extremely fast-moving objects due to the reliance on brief temporal intervals for image reconstruction. A diagram of our solution is described in figure \ref{setup_diagram}.

Our contributions are as follows:
\begin{itemize}
    \item Develop a novel Multispectral imaging reconstruction method.
    \item Collecting labeled dataset of grayscale and MS images across various focal length.
\end{itemize}

\section{Related Work}
Convolutional neural networks (CNNs) have been widely used for image-to-image reconstruction tasks, including depth reconstruction\cite{Mask}, denoising and inpainting\cite{Denoising}, super-resolution\cite{superresolution}, and image colorization\cite{grayscale}. These successes have inspired researchers to apply CNNs to the field of Multispectral (MSI) and hyper-spectral (HSI) imaging.

In MSI, CNNs are often used to solve the ill-posed deconvolution problem, reconstructing a Multispectral cube from a single image or series of images. Hauser and Zeligman\cite{DDNet} used a 2D optical diffuser to create both dispersed and diffused images, which were then reconstructed by a UNet into a hyperspectral cube. However, this method has the drawback of loss in field of view due to replicas on the sensor, as well as a large form factor due to the size of the diffuser.

Jeon et al.\cite{difdractive} introduced a special diffractive optical element that induced a rotated point spread function on a single wavelength, allowing for the reconstruction of a spectral tensor. While this method achieved good results, it was not effective for high spatial frequencies or in the presence of noise.

Oktem et al.\cite{multi} proposed a computational multi-spectral imaging technique using a diffractive lens and algorithms to solve the image reconstruction problem. However, their method was limited to the ultraviolet range and did not use a Fresnel lens like our approach.

Zhao et al.\cite{zhao2022} used a deep learning approach to convert natural color RGB images into high spatial multispectral images, proving particularly useful for precision agriculture and plant phenotyping. However, this approach's applicability may not extend universally across all fields and depends on the quality of the original RGB images. 

In our research,  we utilize a diffractive lens, specifically a Fresnel lens\cite{engelberg2022standardizing}, to disperse and focus light from a scene. Our target is the visible and near-infrared spectrums, and we utilize a UNet-based CNN for the reconstruction of a Multispectral cube. With our method, we intend to address the challenges associated with prior techniques, providing a viable solution to the ill-posed deconvolution issue encountered in MSI.

\section{The Optical System}
\subsection{Fresnel Lens}
A Fresnel lens is a type of lens that works based on diffraction phenomenon using a series of concentric rings. It is thin and lightweight, since it requires significantly less amount of material than a conventional lens. Because of its design, a Fresnel lens can reduce the size and weight of the imaging system, making it easier to deploy in a variety of situations.

\begin{figure}[htbp]
\centering\includegraphics[width=0.5\textwidth]{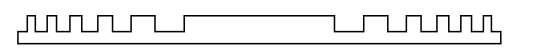}
    \caption{Binary fresnel lens phase profile (Two levels of phase) \cite{DOE}}
    \label{binary_fresnel_lens}
\end{figure}

In this work, we use a binary phase Fresnel Lens, so we focus on this case in our analysis (see figure \ref{binary_fresnel_lens}). We model the wavefront phase as follows:
\begin{equation}
\Delta \phi = \frac{2\pi}{\lambda} n_0 h
\end{equation}
While h represents the height of the element and $n_0$ the refractive index of the material. To enforce the element lens to focus the incident light, it can be shown \cite{video} that there is a relation between the focal lengths $f, f'$, of two wavelength $\lambda, \lambda'$:
\begin{equation}
\frac{f'}{f} = \frac{m\lambda}{m'\lambda'} 
\label{fresnel_f_ratio}
\end{equation}
While $m$ and $m'$ are two integers.

\subsection{The Imaging System}
We denote $u(t)$ as the object's distance from the lens, $z(t)$ as the image sensor distance from the lens and $f(t)$ as the lens' focal distance versus time. The time dependence is derived from the fact that the lens moves in a pre-defined interval $[z_0,z_1]$. We assume that the motion of the lens is small with respect to the distance of the object to the lens:

Assuming the imaging condition is met, we can determine that:
\begin{equation}
\frac{1}{v(t)}+\frac{1}{u(t)}=\frac{1}{f(t)}
\hspace{2cm}
\frac{1}{v(t')}+\frac{1}{u(t')}=\frac{1}{f(t')}
\end{equation}
Merging with equation \ref{fresnel_f_ratio} we get:
\begin{equation}
\frac{z(t)+u(t)}{z(t)u(t)}\frac{z(t')u(t')}{z(t')+u(t')}=\frac{m\lambda(t)}{m'\lambda(t')}
\end{equation}
After some algebra we get:
\begin{equation}
\frac{1}{m'\lambda'(t)}=\frac{1}{m\lambda(t)} \frac{\left[z(t)+u(t)\right]z(t')u(t')}{\left[z(t')+u(t')\right]z(t)u(t)}
\end{equation}
We assume that $u(t)\approx u(t')$ because the object does not move for tiny time intervals. We also assume that the object is far enough so that $u(t)>>v(t),v(t')$ and the motion of the lens is small enough, such that $m'=m=1$. 

Using these approximations, we get the following relation:
\begin{equation}
\frac{1}{\lambda(t)}=\frac{z(t)}{z_0\lambda_0}
\hspace{1cm} \text{or} \hspace{1cm}
\nu(t) = \frac{\nu_0}{z_0} z(t)
\end{equation}
As a result of the inverse relationship between frequency and wavelength for a light wave, a similar relationship can be derived for the frequency $\nu$ of the light.

Our findings indicate that making tiny adjustments to the lens's position is equivalent to focusing on different light frequencies. However, it is important to note that, for each distinct lens distance, only one specific wavelength is sharply in focus. While the image still captures other wavelengths, they appear out of focus.  As such, we concentrate our efforts on the image edges, which are more prone to blurriness. The reasoning behind this approach will be further elaborated on later in this paper.

\section{Deep Neural Network Model}
\subsection{Architecture}
In our proposed model, we based on UNet architecture \cite{UNET} and used PyTorch implementation. This architecture was chosen due to its known ability to effectively preserve fine details, which is essential for accurate Multispectral imaging \cite{macdonald2017assessment}. The UNet structure is specifically designed to facilitate the flow of information from the encoder to the decoder, while also enabling skip connections that fuse features from different levels of abstraction. This proves particularly advantageous for our image-to-image translation task, as it allows for the recovery of fine details from the input images and the subsequent generation of a high-resolution output image.

The model's input size is $C_{in} \times 256 \times 256$, while $C_{in}$ represents the number of channels in the input. Similarly, the output size, for each image is $C_{out} \times 256 \times 256$.
We experimented with different input and output channel combinations but primarily focused on settings where $C_{in}=50$ and $C_{out}=50$. This choice was guided by resources versus performance trade-off considerations.

To increase the diversity of the training data and enhance the model's generalization capabilities, we applied data augmentations in our training process. These augmentations include transformations such as translation, rotation, cropping, and flipping.
\subsection{Data Preprocess}

After capturing the sequence of grayscale images, we undertook three preprocessing steps:
\begin{enumerate}
    \item As the change in lens distance from the sensor results in differing imaging magnification, an alignment process between the images is necessary. To achieve this, we employed an Affine transformation, which maps all images to the same coordinate system to ensure their complete alignment with each other. This transformation was based on an automatic corner detector and point correspondence. 
    \item As previously explained, to encourage the model focus on differences between images, we used a Sobel Edge detector on each of the image sequences. This approach was justified by the observation that sharp focus or severe defocus is particularly distinguishable near the edges.
    \item Then, to encourage the model to focus on the differences between wavelengths, we transformed the image sequence into a temporal differences sequence, which is equivalent to taking the temporal derivative.
\end{enumerate}

\subsection{The Loss}
Our supervised model was labeled pixel-wise by the MS cube. However, we added two regularization based on prior knowledge. We used the Total Variation - for the Spatial and Spectral domain together:
\begin{equation}
\sum_{i,j,k} \sqrt{(\Delta_h \hat{y}_{i,j,k})^2 + (\Delta_v \hat{y}_{i,j,k})^2+\gamma_{TVS}(\Delta_k \hat{y}_{i,j,k})^2 }
\end{equation}
While $\gamma_{TVS}$ is a hyperparameter that describes the un-isotropic rate of the TV loss between the spectral and spatial axes.
The TV loss measures the variation in pixel intensities in the input image by computing the magnitude of the gradients in the horizontal, vertical and spectral directions \cite{chan2006total}. The loss is then calculated as the sum of the magnitude of the gradients across all pixels in the image. And the Structural Similarity Measurement after reconstruction of the Multispectral images, we project them on RGB basis, and then we used SSIM \cite{1284395}.

Finally, we used a combination of the terms as follows:
\begin{equation}
\mathcal{L}(y,\hat{y}) = \left\| y - \hat{y} \right\|_1 + \lambda_{TV} L_{TV}(y,\hat{y}) + \lambda_{SSIM}\left(1-SSIM(y,\hat{y})\right)
\end{equation}
While $y$ is the label, $\hat{y}$ is the model outputs and  $\lambda_{TV}$, $\gamma_{TVS}$, and $\lambda_{SSIM}$ are hyperparamets. We empirically found that setting them to $0.1, 0.2, 0.9$, respectively, works well.

The combined loss function allows the model to consider multiple aspects of image quality, including pixel-wise differences, structural similarity, and the smoothness of the output image.

\section{Experiment}
\subsection{Setup}
\begin{figure}[htbp]
\centering\includegraphics[width=1.0\textwidth]{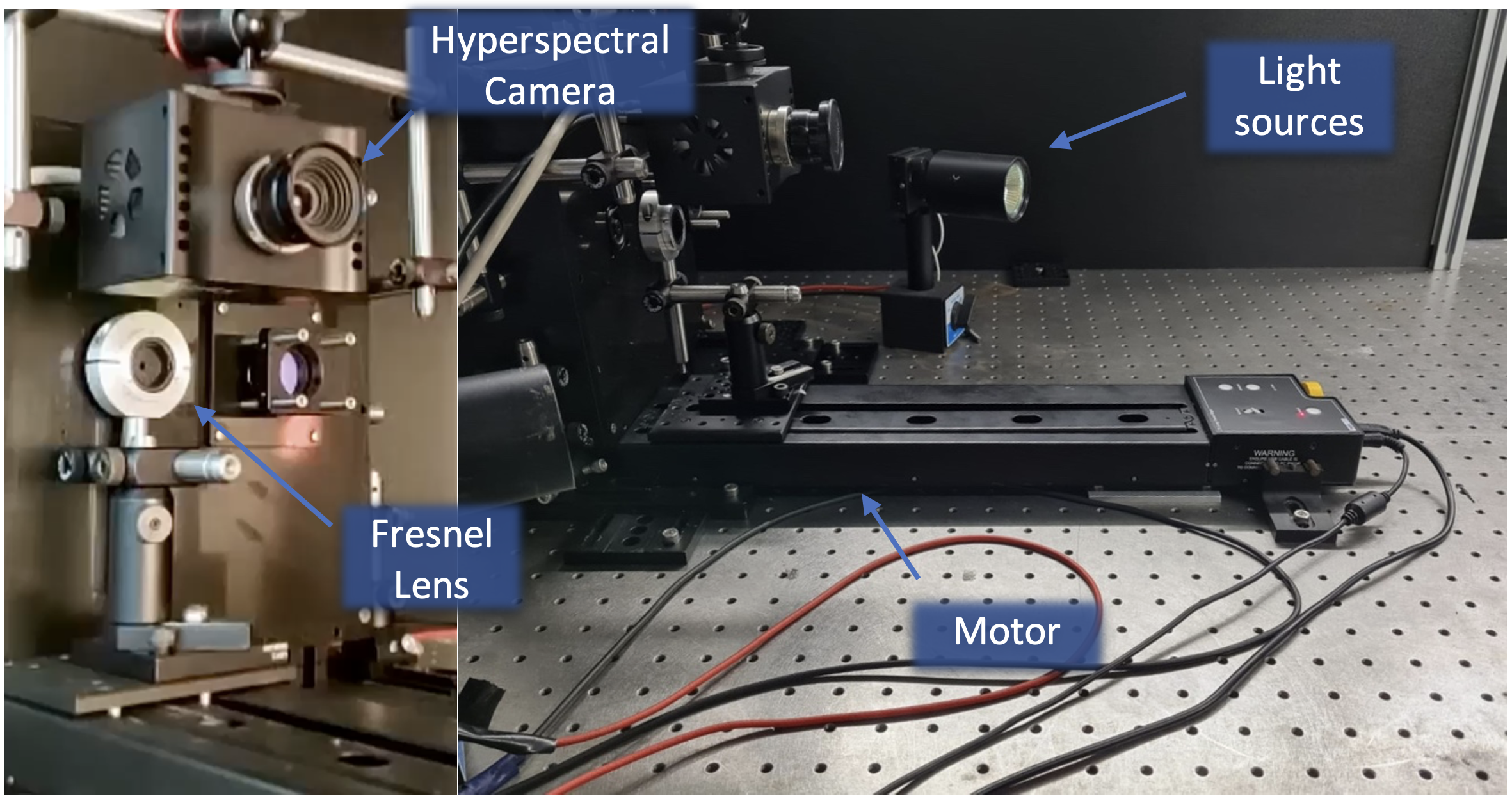}
\caption{Camera setup}
\label{experimental_setup}
\end{figure}

Our system consists of an optical set-up for data acquisition and an algorithm layer for reconstruction of the Multispectral (MS) cube. The optical set-up, as can be seen in figure \ref{experimental_setup}, includes a motor stage (Thor-Labs 300mm stage), a Fresnel lens, and a gray-scale sensor (Onsemi MT9P031).
 The motor stage is intended to function similarly to an optical auto-focus mechanism by changing the distance of the lens from the sensor, which in turn changes the focus pattern that is learned by the neural network. The Fresnel lens is diffractive and disperses color along the spatial domain, and its Modulation Transfer Function (MTF) and Field of View (FOV) was measured and attached in figures \ref{fov_measurement},\ref{mtf_measurement} in Appendix \ref{appendix}. The gray-scale sensor, with a pixel size of $2.5 \times 2.5$ microns and $2592 \times 1944$ active pixels \cite{nesemi}. Additionally, an imec MSI snap-scan VNIR camera \cite{imec} is used to capture the ground-truth for the supervised training. The set-up was fully automated using python APIs of each device to perform the acquisition, including capturing with the gray-scale sensor, moving the motor, and finally acquiring the MS cube as a ground truth.

\subsection{Labeling Mapping}

The point of view of our system's camera and the label data (MS camera) is different, so to map the MS cube to the model's output MS cube's coordinate, we used homography transformation between the images.

\begin{figure}[htbp]
\centering\includegraphics[width=1.0\textwidth]{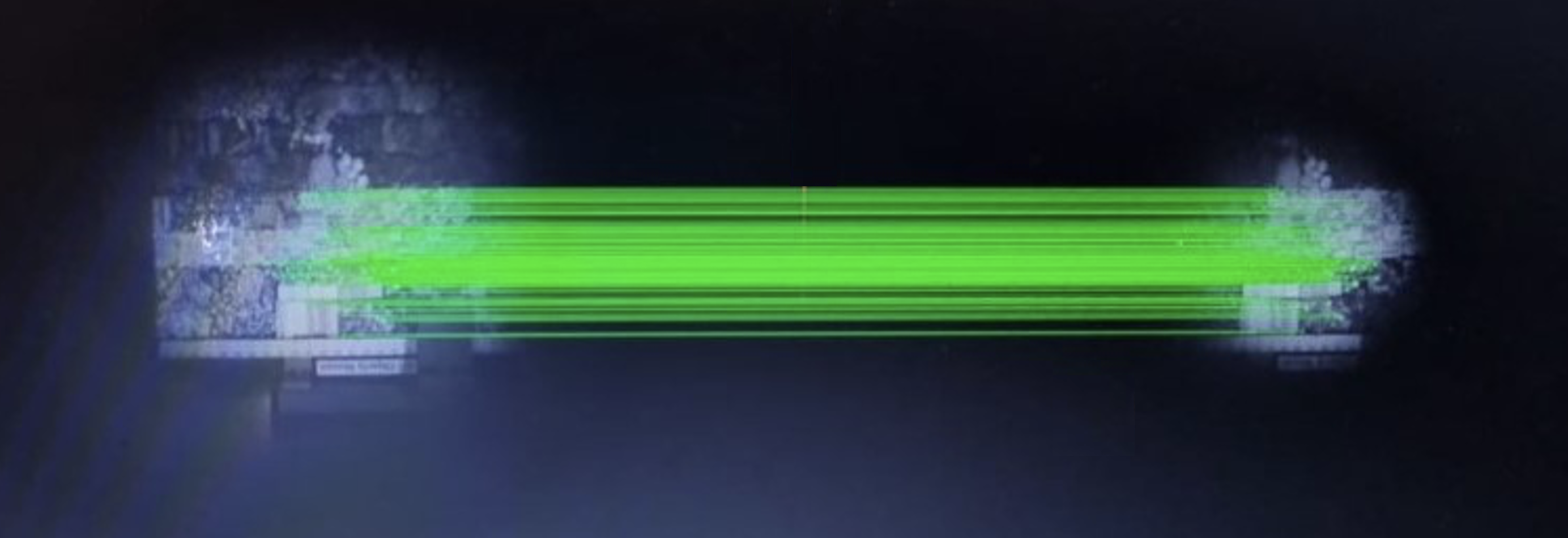}
\caption{Mapping from the Multispectral camera to our camera with Homography transformation using automatic detection and matching}
\label{homography_example}
\end{figure}

Our homography transformation is illustrated in Figure \ref{homography_example}. We used the ORB feature detector\cite{6126544} and Brute Force Matcher to compare the input and label data, using Hamming Distance as the matching criterion. 
The recorded objects was roughly at a constant distance from the camera (1.3 meters) and placed on a flat board. This approach was selected to ensure that the scenes were flat, as the selected homography technique is not designed to accommodate the fitting of three-dimensional objects. With the scene properly aligned, our automated system algorithm controlled and synchronized between the motor and the camera.

The dataset contains measurements of 52 objects we recorded in our lab, using halogen light bolbs illumination. Each object was recorded by the Multispectral camera (the ground truth) in 150 spectral channels within the wavelength range of [470 nm , 900 nm] and series of gray scale images captured for different lens distance.

\section{Results}
\begin{figure}[htbp]
\centering\includegraphics[width=1\textwidth]{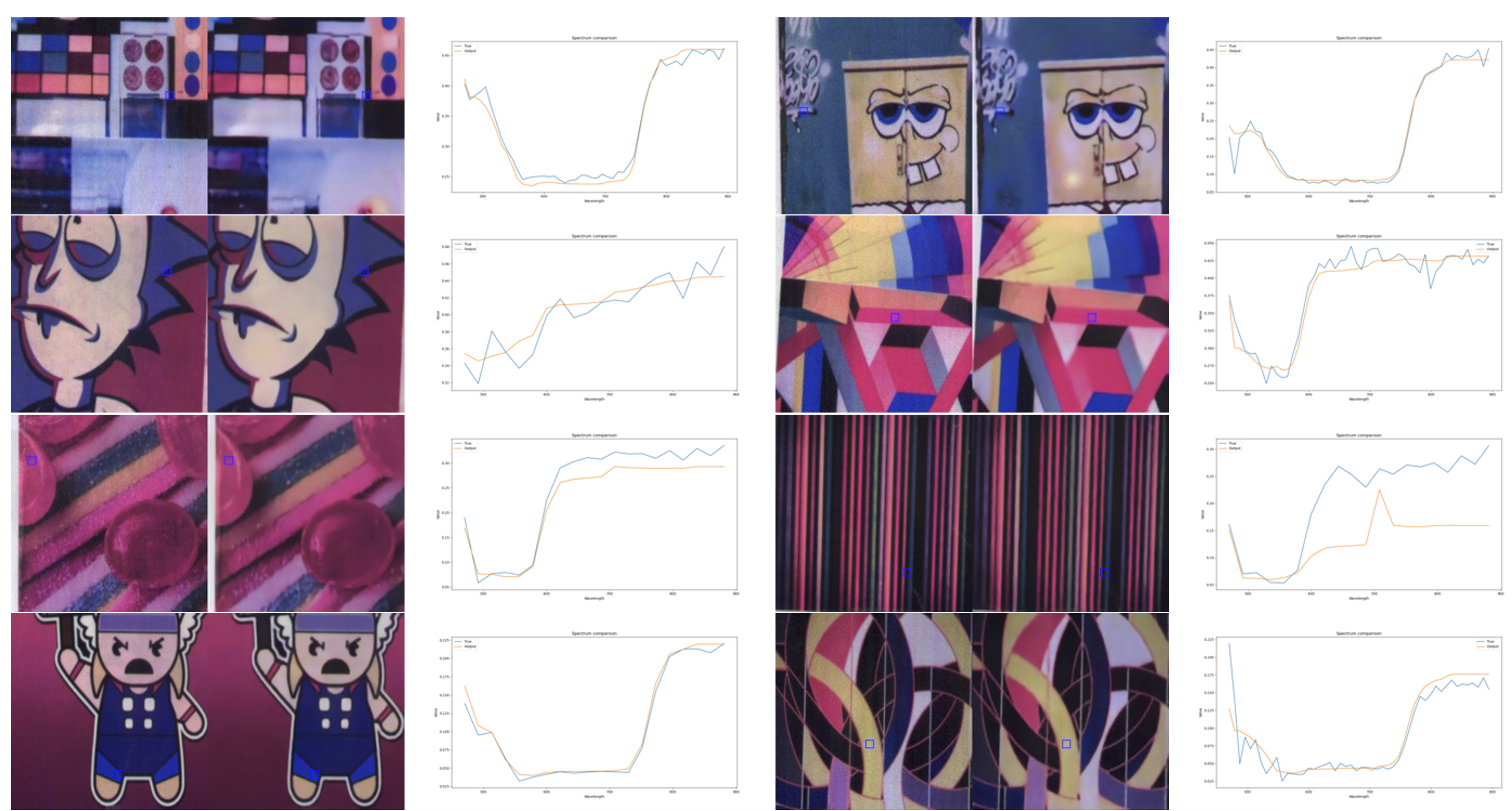}
\centering\includegraphics[width=1\textwidth]{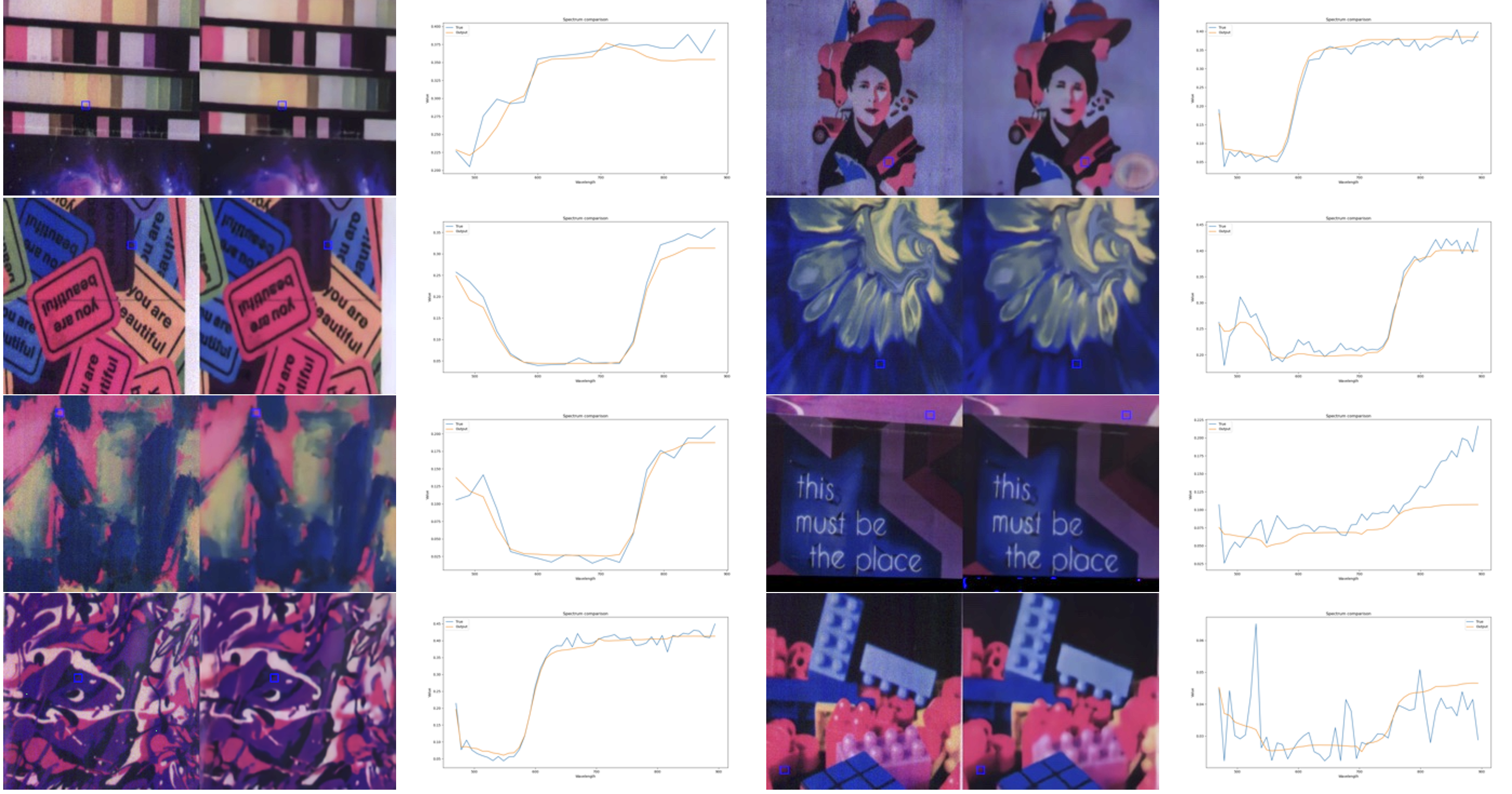}
\caption{Image reconstruction ground truth (left) versus our camera (right). The plots display a random pixel Multispectral signature comparison between the ground truth (blue) and our model (orange)}
\label{reconstructions_examples}
\end{figure}

\begin{figure}[htbp]
\centering\includegraphics[width=1\textwidth]{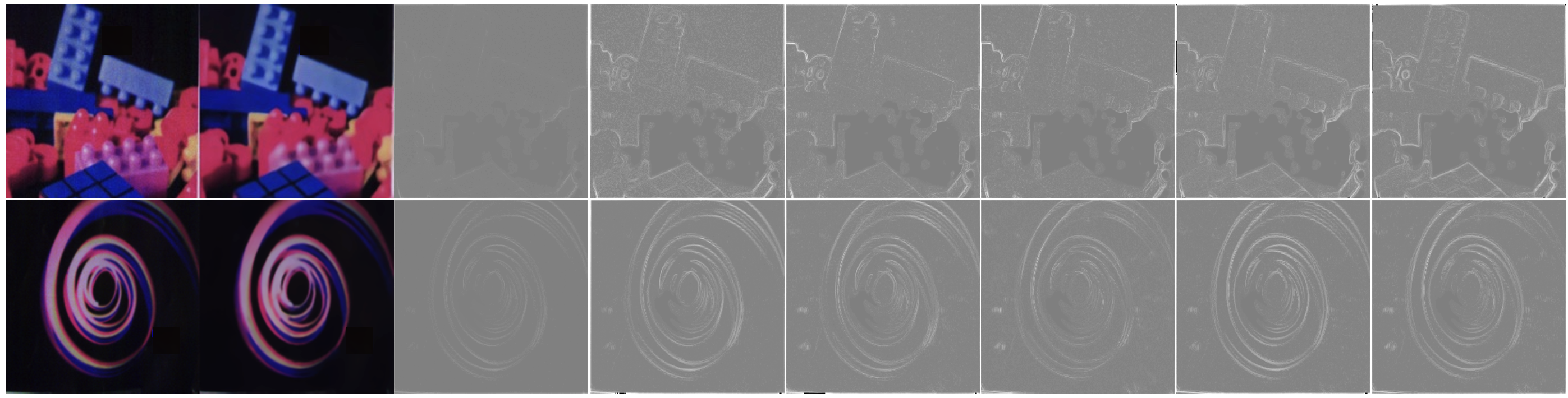}
\caption{Input sequence to the model and its output (left), the model output versus ground truth (right)}
\label{input_sequence}
\end{figure}

Some of the results are presented in figure \ref{reconstructions_examples}.
An illustration of the input and output of the model is shown in figure \ref{input_sequence}.

The concept of our fresnel lens MSI camera was proven for \( L = 50 \) wavelength channels in a range from 470 nm to 900 nm. The latest outcomes reveal notable advancements in our network's proficiency to learn and reconstruct the spatial and spectral characteristics of objects as can be seen in figure \ref{reconstructions_examples}. 
To asses out system performance we used 11 test cubes that were isolated from the training set. We used 3 metrics, PSNR, SSIM and L1 for quality evaluation as can be seen in table\ref{Experimental_results}.
Nonetheless, we recognize that there is room for further improvement. As depicted in the bottom right of figure, we present an object that the net was trained on, clearly demonstrating that the training process is still underway. We remain confident that by collecting a more extensive and diverse training data, our network will acquire a more comprehensive grasp of various scenarios, leading to heightened performance in reconstruction tasks. With these measures in place, we anticipate significant progress in our network's overall generalization and proficiency.

\begin{table*}[ht!]
\begin{center}
\begin{tabular}{ c | c | c | c }
 & Best & Worst & Mean \\ \hline
PSNR (DB) & 24.267 & 23.873 & 24.058 \\  
SSIM & 0.723 & 0.690 & 0.722 \\
L1 & 0.041 & 0.050 & 0.042 \\
\end{tabular}
\end{center}
\caption{Experimental results.}
\label{Experimental_results}
\end{table*}

\subsection{Compare with similar methods}
Can be seen in table\ref{tab:my_table}.
\begin{table*}[ht!]
\begin{center}
 \begin{tabular}{ c | c | c| c | c }
  & DD-Net \cite{DDNet} & MS\cite{multi} & DR\cite{difdractive} & Ours \\ \hline
 DOE in use & Diffuser & Photon sieve & Self-fab & Fresnel Lens \\  
Spatial Resolution & $256\times256$ & Diff Lim & $256\times256$ & $256\times256$ \\
Color Channels Number & 29 & ~4 & 25 & 50 \\
 Spectral region of interest & VIS & EUV & VIS & VIS \\
 \end{tabular}
 \end{center}
  \caption{Comparison with similar methods.}
  \label{tab:my_table}
 \end{table*}

\newpage
\section{Discussion: Potential Areas for Improvement and Future Work}
In our study, we identified key aspects that provide opportunities for refining our methodology. Our distances from the sensor v, were not chosen according to the dependency of the focused wave-length on v, but arbitrarily. This arbitrary selection limits our ability to ensure the entire region of interest in the spectrum is in focus during grayscale image capture. 

 Attention to potential error sources, such as motor inaccuracies and the monochrome detector's dynamic range, can be addressed through precise step-size selection and SNR management. 

The exploration of alternative diffractive lens options and targeted strategies to address FOV limitations present exciting avenues for improving the system's performance. Furthermore, expanding our algorithm to accommodate 3D object reconstruction is another promising direction for future work.

\section{Conclusion}
In this work, we have demonstrated a novel approach to Multispectral imaging using an innovative combination of a Fresnel lens-based optical setup and a deep learning model. Our use of a U-Net architecture, which is known for preserving fine details in imaging tasks, enabled us to extract and process Multispectral information from grayscale images. Through sophisticated data preprocessing and augmentation methods, we ensured high-quality training data for our model.

Despite the challenges encountered during the implementation, the model has shown promising results in reconstructing the spatial and spectral information of various objects. The limitations in our current approach, such as inaccuracies in color reconstruction and difficulties with 3D object representation, provide us with opportunities for further research and improvement.

Notwithstanding these limitations, we believe that our method holds great potential for providing a cost-effective and efficient solution for Multispectral imaging. With further optimization and training of the model, we anticipate improved performance and applicability to a wide range of practical scenarios.

Our research thus contributes to the broader field of Multispectral imaging, illustrating the potential of integrating traditional optical systems with advanced machine learning techniques. Going forward, we hope to refine our approach and explore its potential for real-world applications.

\newpage
\newpage
\bibliographystyle{unsrtnat}
\bibliography{main}  %%% Uncomment this line and comment out the ``thebibliography'' section below to use the external .bib file (using bibtex) .

\section{Appendix}
\label{appendix}

\subsection{Binary Fresnel Lens}
\begin{figure}[htbp]
\centering\includegraphics[width=1.\textwidth]{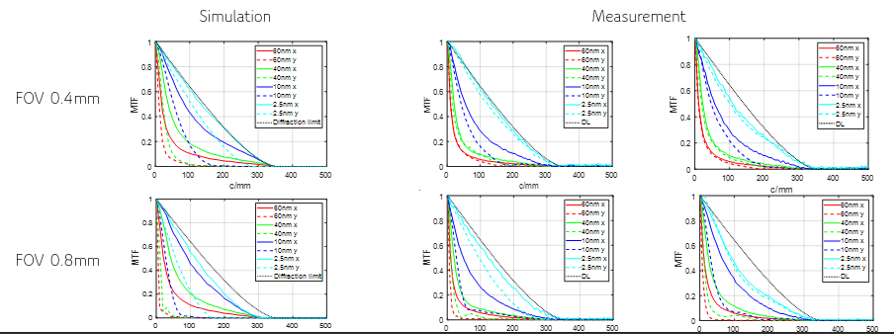}
\caption{FOV measurements of our Binary Fresnel Lens}
\label{fov_measurement}
\end{figure}

\begin{figure}[htbp]
\centering\includegraphics[width=1\textwidth]{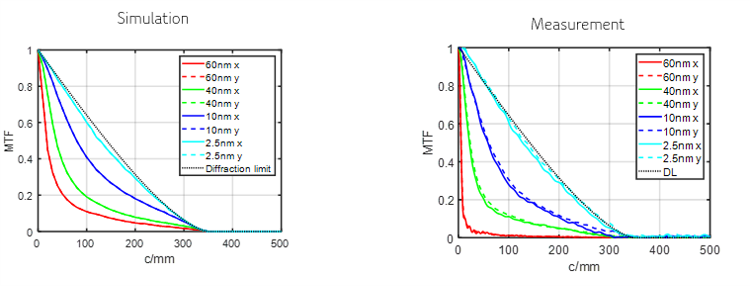}
\caption{MTF measurements of our Binary Fresnel Lens}
\label{mtf_measurement}
\end{figure}

\newpage
\section*{Backmatter}
    % \section*{Funding}
    % Content in the funding section will be generated entirely from details submitted to Prism. Authors may add placeholder text in the manuscript to assess length, but any text added to this section in the manuscript will be replaced during production and will display official funder names along with any grant numbers provided. If additional details about a funder are required, they may be added to the Acknowledgments, even if this duplicates information in the funding section. See the example below in Acknowledgements.

    \subsection*{Acknowledgments}
    We would like to thank Dr. Jacobb Engelbert and Prof. Uriel Levy for creating the Fresnel Lens. In addition, We would like to thank Adir Cohen and Ido Keynan for their help in building the Setup.

    \subsection*{Disclosures}
    The authors declare no conflicts of interest.

    \subsection*{Data Availability Statement}
    Data underlying the results presented in this paper are available in \cite{data}.
\end{document}